\documentclass[11pt,draftcls,onecolumn]{IEEEtran}

\IEEEoverridecommandlockouts

\usepackage[cmex10]{amsmath}
\usepackage{amssymb}

\usepackage{graphicx}

\newtheorem{defi}{Definition}
\newtheorem{theorem}{Theorem}
\newtheorem{lemma}{Lemma}
\newtheorem{remark}{Remark}

\newcommand{\Xc}{\mathcal{X}}
\newcommand{\Yc}{\mathcal{Y}}

\newcommand{\Cc}{\mathcal{C}}
\newcommand{\Mc}{\mathcal{M}}

\newcommand{\Rc}{\mathcal{R}}

\newcommand{\Nc}{\mathcal{N}}
\newcommand{\Pc}{\mathcal{P}}

\begin{document}
%
\title{Three-User Cognitive Interference Channel: Capacity Region with Strong Interference}
%
%

\author
{
Mahtab Mirmohseni, Bahareh Akhbari, and Mohammad Reza Aref\\
Information Systems and Security Lab (ISSL)\\
Department of Electrical Engineering, Sharif University of Technology, Tehran, Iran \\
Email: mirmohseni@ee.sharif.edu, b\_akhbari@ee.sharif.edu, and
aref@sharif.edu %
\thanks{This work was partially supported by Iran National Science Foundation (INSF) under contract No. 88114.46-2010 and by Iran Telecom Research Center (ITRC) under contract No. T500/17865.}}

\maketitle

\begin{abstract}
 This study investigates the capacity region of a three-user cognitive radio network with two primary users and one cognitive user. A  three-user Cognitive Interference Channel (C-IFC) is proposed by considering a three-user Interference Channel (IFC) where one of the transmitters has cognitive capabilities and knows the messages of the other two transmitters in a non-causal manner. First, two inner bounds on the capacity region of the three-user C-IFC are obtained based on using the schemes which allow all receivers to decode all messages with two different orders. Next, two sets of conditions are derived, under which the capacity region of the proposed model coincides with the capacity region of a three-user C-IFC in which all three messages are required at all receivers. Under these conditions, referred to as strong interference conditions, the capacity regions for the proposed three-user C-IFC are characterized. Moreover, the Gaussian three-user C-IFC is considered and the capacity results are derived for the Gaussian case. Some numerical examples are also provided.
\end{abstract}



%
\IEEEpeerreviewmaketitle


\section{Introduction}
%
%
%
%

Interference avoidance techniques have traditionally been used in wireless networks wherein multiple source-destination pairs share the same medium. However, the broadcasting nature of wireless networks may enable cooperation among entities, which ensures higher rates with more reliable communication. On the other hand, due to the increasing number of wireless systems, spectrum resources have become scarce and expensive. The exponentially growing demand for wireless services along with the rapid advancements in wireless technology have lead to cognitive radio technology which aims to overcome the spectrum inefficiency problem by developing communication systems that have the capability to sense the environment and adapt to it \cite{GolJafMarSri09,Mit91}.

In overlay cognitive networks, the cognitive user can transmit simultaneously with the non-cognitive users and compensate for the interference by cooperation in sending, i.e., relaying, the non-cognitive users' messages \cite{GolJafMarSri09}. In order to obtain the fundamental limits of these networks by information theoretical techniques, researchers have to consider the models with idealized assumptions. The assumption of \emph{full non-causal} knowledge of the primary messages (as side information) at the cognitive users is a standard one, which is still very useful in practical applications if one considers a phase for obtaining this side information. From an information theoretic point of view, Cognitive Interference Channel (C-IFC) was first introduced in \cite{DevrMitTar06} to model an overlay cognitive radio and refers to a two-user Interference Channel (IFC) in which the cognitive user (secondary user) has the ability to obtain the message being transmitted by the other user (primary user), either in a non-causal or causal manner. For the non-causal C-IFC, where the cognitive user has non-causal full or partial knowledge of the primary user's transmitted message, an achievable rate region was first derived in \cite{DevrMitTar06}, by combining the Gel'fand-Pinsker (GP) binning \cite{GelfPin80} with a well known simultaneous superposition coding scheme (rate splitting) applied to IFC \cite{HanKob81}. Subsequently, several achievable rate regions and capacity results in some special cases for the C-IFC have been established \cite{JoviVis09}-\cite{JiangXin08}. Yet, capacity results have been known only in special cases. C-IFC with strong interference conditions is one of these cases, where interference is such that both messages can be decoded at both receivers with no rate penalty. Strong interference conditions for C-IFC and the capacity regions under these conditions have been derived in \cite{MariYatKra07, MirAkhArefITW10, MariYatKra06}. For an overview on the capacity results of C-IFC, see \cite{RiniIT11}.

The \emph{$k$-user} IFC consists of $k$ independent transmitters sending messages to $k$ independent receivers. Extending the results of the classic two-user IFC to the IFCs with more than two user pairs is non-trivial; because each receiver is affected by the joint interference from the all other transmitters rather by each transmitter's signal separately \cite[P.~157]{ElgKim11}. Recently, the capacity region of a three-user Gaussian IFC under mixed strong-very strong interference conditions has been characterized in \cite{ChaSez10}. In C-IFC, asymmetric nature of the transmitters' cooperation makes this extension even more challenging, since there are several ways for applying the cognition capabilities and also the obtained setups may involve different aspects of IFCs such as independent channel inputs at the transmitters which makes difficult to apply the results of C-IFC to these setups. An achievable rate region for a three-user Multiple Access Channel (MAC) \cite[Chapter~4]{ElgKim11} with three transmitters and \emph{one receiver} has been derived in \cite{NagKis_Gl10}. By increasing the number of receivers, a \emph{three-user} C-IFC with \emph{one} primary user and \emph{two} cognitive users has been studied in \cite{NagMur_ITW09, NagMur_Gl09}, where an achievable rate region is derived for this setup. The authors in \cite{NagMohMurKis_11}, proposed the achievable rate regions for the different non-causal message-sharing mechanism in the three-user C-IFC and also derived an outer bound in the Gaussian case.

In this paper, we consider a \emph{three-user} C-IFC with \emph{two} primary users and \emph{one} cognitive user, where the cognitive transmitter non-causally knows the messages of both primary transmitters. Up to our best knowledge, in all of the previous works on \emph{three-user} C-IFC in the general discrete memoryless setup, only achievable rate regions have been obtained and the capacity result in all setups of \emph{three-user} C-IFC is an open problem. In this paper, we consider the strong interference regime and derive \emph{capacity regions} in this case. First, we obtain two inner bounds on the capacity region (achievable rate regions) based on using superposition coding and allowing all receivers to decode all messages. In the achievablity scheme of the first region, we utilize simultaneous joint decoding in the decoding part at all receivers.
However, in the second scheme, each primary receiver first decodes the other primary user's message, while treating the remaining signals as noise, i.e., the combination of its intended transmitter's signal, the cognitive transmitter's signal and additive noise. This strategy is useful for the channels where the other primary user's signal (as seen by each primary user) is strong enough and it is possible to decode this primary interference first. Then, the primary receiver decodes the message of the cognitive user and its own message by a joint typicality decoding. The receiver of the cognitive user pair uses joint typicality decoding. Next, deriving two sets of strong interference conditions, we show that the obtained inner bounds achieve capacity under these conditions by proving converse proofs. In these cases, decoding the unintended messages causes no additional constraint on the rate region. Therefore, the channel model is equal to the one in which all three messages are required at all receivers and the capacity region coincides with the capacity region of a three-user C-IFC in which each receiver should decode all three messages. In fact, we determine the conditions, referred to as \textit{Set1}, under which the three-user C-IFC can be seen as a compound three-user MAC with common information. Under the second set of conditions, referred to as \textit{Set2}, the considered channel can be seen as a compound of three channels: two \textit{two-user MAC}s with common information at the primary receivers and a three-user MAC with common information at the cognitive receiver. Further, we compare these two sets of conditions and show that \textit{Set1} is weaker than \textit{Set2}. Moreover, we consider the Gaussian three-user C-IFC and find capacity results for the Gaussian case based on \textit{Set2}. We also provide some numerical examples.

The rest of the paper is organized as follows. Section~\ref{sec:definition} introduces the three-user C-IFC model and the notations. In Section~\ref{sec:Ach}, we obtain the achievable rate regions; while in Section~\ref{sec:Cap}, we state the capacity results for the discrete memoryless three-user C-IFC. In Section~\ref{sec:Gaussian}, Gaussian three-user C-IFC is investigated. Finally, Section~\ref{sec:conclusion} concludes the paper.

\section{Channel Models and Preliminaries}\label{sec:definition}
Throughout the paper, upper case letters (e.g. $X$) are used to denote RVs and lower case letters (e.g. $x$) show their realizations. The probability mass function (p.m.f) of a RV $X$ with alphabet set $\Xc$, is denoted by $p_X(x)$, where subscript $X$ is occasionally omitted. $A_\epsilon^n(X,Y)$ specifies the set of $\epsilon$-strongly, jointly typical sequences of length $n$. The notation $X^j_i$ indicates a sequence of RVs $(X_i,X_{i+1},...,X_j)$, where $X^j$ is used instead of $X^j_1$, for brevity. $\Nc(0,\sigma^2)$ denotes a zero mean normal distribution with variance $\sigma^2$.

\begin{figure}[tb]
  \centering
  \includegraphics[width=11cm]{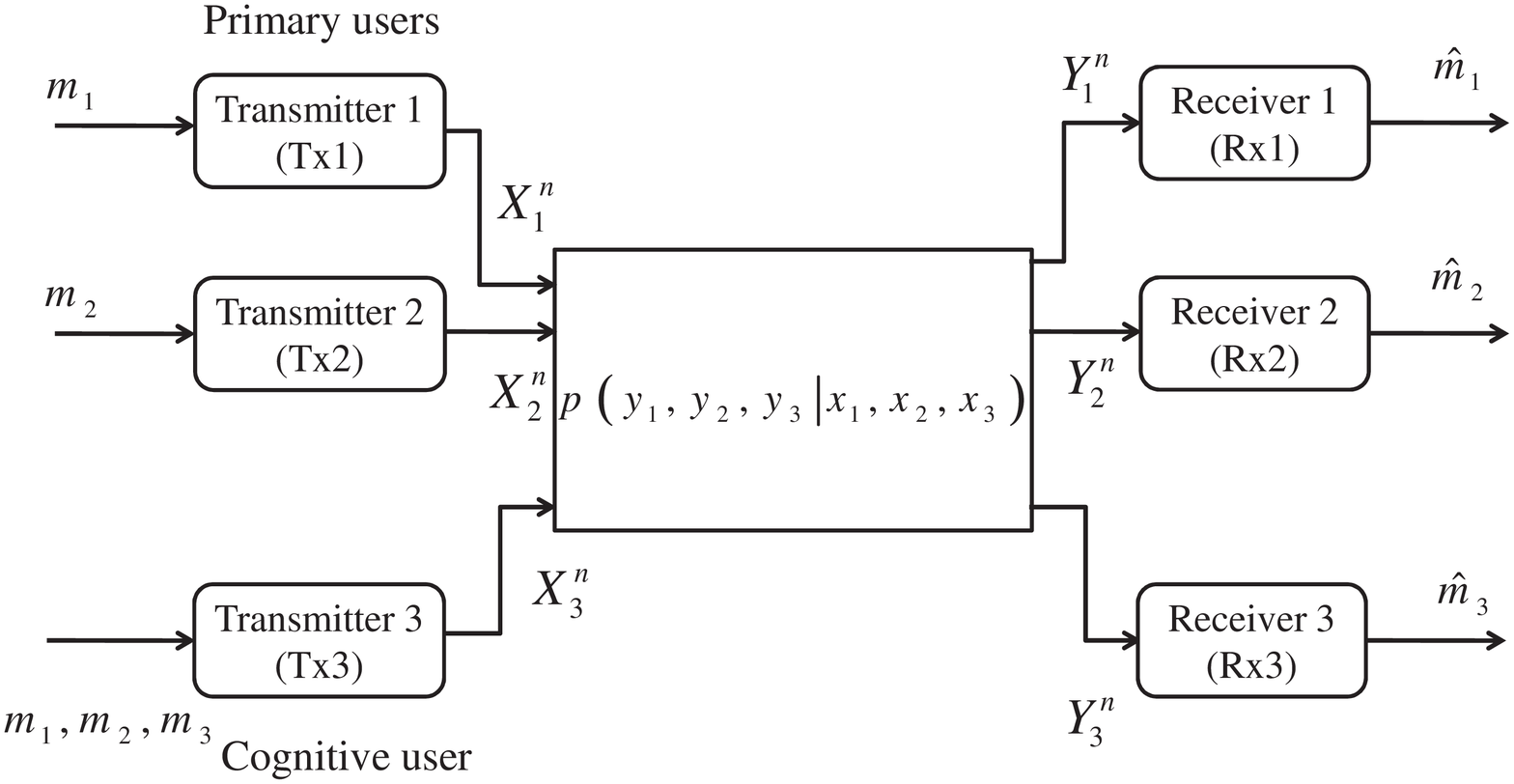}
  \caption{Three-user Cognitive Interference Channel (C-IFC)}
  \label{fig:channelmodel}
\end{figure}
Consider the three-user C-IFC in Fig.\ref{fig:channelmodel}, which is denoted by ($\Xc_1\times\Xc_2\times\Xc_3,p(y_1^n,y_2^n,y_3^n|x_1^n,x_2^n,x_3^n),\Yc_1\times\Yc_2\times\Yc_3$), where $X_u\in\Xc_u$ is the channel input of Transmitter~$u$ (Tx$u$) and $Y_u\in\Yc_u$ is the channel output at Receiver~$u$ (Rx$u$) for $u\in\{1,2,3\}$. Also, $p(y_1^n,y_2^n,y_3^n|x_1^n,x_2^n,x_3^n)$ is the channel transition probability distribution. In $n$ channel uses, each Tx$u$ desires to send a message $m_u$ to Rx$u$ where $u\in\{1,2,3\}$.

\begin{defi}\label{def:code}
A $(2^{nR_1},2^{nR_2},2^{nR_3},n)$ code for the three-user C-IFC consists of (i) three independent message sets $\Mc_u=\{1,...,2^{nR_u}\}$, where $u\in\{1,2,3\}$, (ii) two encoding functions at the primary transmitters, $f_1:\Mc_1\mapsto\Xc_1^n$ at Tx1 and $f_2:\Mc_2\mapsto\Xc_2^n$ at Tx2, (iii) an encoding function at the cognitive transmitter, $f_3:\Mc_1\times\Mc_2\times\Mc_3\mapsto\Xc_3^n$, and (iv)~three decoding functions, $g_u:\Yc_u^n\mapsto\Mc_u$ at Rx$u$ where $u\in\{1,2,3\}$. We assume that the channel is memoryless. Thus, the channel transition probability
distribution is given by
\begin{equation}\label{eqn:pmf}
p(y_1^n,y_2^n,\,y_3^n|x_1^n,x_2^n,x_3^n)=\prod\limits_{i=1}^np(y_{1,i},y_{2,i},y_{3,i}|x_{1,i},x_{2,i},x_{3,i}).
\end{equation}

The probability of error for this code is defined as $P_e=max\{P_{e,1},P_{e,2},P_{e,3}\}$, where we have
\begin{IEEEeqnarray*}{l}
P_{e,u}=\frac{1}{2^{n(R_1+R_2+R_3)}}\sum\limits_{m_1,m_2,m_3}{P(g_u(Y^n_{u})\neq m_u | (m_1,m_2,m_3)\textrm{ sent})}
\end{IEEEeqnarray*}
for $u\in\{1,2,3\}$.
\end{defi}

\begin{defi}\label{def:rate}
A rate triple $(R_1,R_2,R_3)$ is achievable if there exists a sequence of $(2^{nR_1},2^{nR_2},2^{nR_3},n)$ codes with $P_e\rightarrow 0$ as $n\rightarrow \infty$. The capacity region $\Cc$, is the closure of the set of all achievable rates.
\end{defi}

\section{Achievable Rate Regions for Discrete Memoryless three-user C-IFC}\label{sec:Ach}
In this section, we consider the discrete memoryless three-user C-IFC and present two achievable rate regions for this setup. The coding schemes contain superposition coding in the encoding part. In the decoding part, all messages are common to all receivers, i.e., all three receivers decode $m_1$, $m_2$ and $m_3$. In the scheme of the first achievable rate region, the simultaneous joint decoding is utilized at all receivers. However, in the second scheme, Rx1 first decodes the other primary user's message $m_2$, while treating the remaining signals as noise, i.e., the signals of $m_1$ and $m_3$ plus additive noise. This strategy is useful for the channels where the signal of $m_2$ at Rx1 is strong enough and it is possible to decode this primary interference first. Then, Rx1 decodes the message of the cognitive user $m_3$ and its own message $m_1$ by a joint typicality decoding. Rx2 proceeds similarly, while, Rx3 uses joint typicality decoding. Detailed proofs are provided in Appendix~\ref{app:ach_proof}.

Let $\Pc$ denotes the set of all joint p.m.fs $p(.)$, that factor as 
\begin{IEEEeqnarray}{c}
p(x_1,x_2,x_3)=p(x_1)p(x_2)p(x_3|x_1,x_2).\label{eqn:pmf_ach}
\end{IEEEeqnarray}

\begin{theorem}\label{thm:ach1}
The union of rate regions given by
\begin{IEEEeqnarray}{rcl}
R_{3}&\leq& I(X_3;Y_3|X_1,X_2) \label{eqn:ach1_I}\\
R_{1}+R_{3}&\leq& \min\{I(X_1,X_3;Y_1|X_2),I(X_1,X_3;Y_3|X_2)\} \label{eqn:ach1_II}\\
R_{2}+R_{3}&\leq& \min\{I(X_2,X_3;Y_2|X_1),I(X_2,X_3;Y_3|X_1)\} \label{eqn:ach1_III}\\
R_{1}+R_{2}+R_{3}&\leq&\min\{I(X_1,X_2,X_3;Y_1),I(X_1,X_2,X_3;Y_2),I(X_1,X_2,X_3;Y_3)\} \label{eqn:ach1_IV}
\end{IEEEeqnarray}
is achievable for the three-user C-IFC (denoted as $\Rc_1(p)$), where the union is over $p(.)\in\Pc$ (defined in (\ref{eqn:pmf_ach})).
\end{theorem}

\begin{theorem}\label{thm:ach2}
The union of rate regions given by (\ref{eqn:ach1_I})-(\ref{eqn:ach1_III}) and
\begin{IEEEeqnarray}{rcl}
R_{1}&\leq& I(X_1;Y_2) \label{eqn:ach2_I}\\
R_{2}&\leq& I(X_2;Y_1)  \label{eqn:ach2_II}\\
R_{1}+R_{2}+R_{3}&\leq& I(X_1,X_2,X_3;Y_3) \label{eqn:ach2_III}
\end{IEEEeqnarray}
is achievable for the three-user C-IFC (denoted as $\Rc_2(p)$), where the union is over $p(.)\in\Pc$ (defined in (\ref{eqn:pmf_ach})).
\end{theorem}


\section{Strong Interference Conditions and Capacity Results}\label{sec:Cap}
In this section, we derive two sets of strong interference conditions (\textit{Set1} and \textit{Set2}), under which the regions of Theorem~\ref{thm:ach1} and Theorem~\ref{thm:ach2} achieve capacity. First, we give an intuition about deriving the conditions at each receiver:
\begin{itemize}
\item \textbf{Strong interference at the cognitive receiver (Rx3):} In \textit{both schemes}, Rx3 jointly decodes $m_1$, $m_2$ and $m_3$. Therefore, it is assumed that $m_1$ and $m_2$ jointly cause strong interference. These conditions are shown in (\ref{eqn:set1_IV}) and the second terms of (\ref{eqn:set1_II}) and (\ref{eqn:set1_III}) for the first scheme. In other words, assuming the above conditions, the joint received signal from Tx1 and Tx2 at Rx3 is strong enough to decode without imposing any rate constraint on $R_{1}$ and $R_{2}$. Similar conditions are also provided for the second scheme in (\ref{eqn:set2_IV}) and the second terms of (\ref{eqn:set2_II}) and (\ref{eqn:set2_III}). Therefore, there is no difference between two schemes about the strong interference condition at the cognitive receiver (Rx3).

\item \textbf{Strong interference at the primary users (Rx1 and Rx2):} We illustrate the condition for Rx1 and the one for Rx2 follows due to the symmetry. In the first scheme, condition at Rx1 is similar to Rx3 and it is assumed that $m_2$ and $m_3$ jointly cause strong interference, which is shown in the first terms of (\ref{eqn:set1_I}) and (\ref{eqn:set1_III}). Note that, the asymmetric nature of the conditions, compared to the one for Rx3, is due to the cognition capability of Tx3, i.e., $x_3$ depends on $m_1$ and $m_2$ in addition to $m_3$. However, in the second scheme, it is assumed that the interference caused by $m_2$ at Rx1 is stronger than the joint received signals of $m_1$ and $m_3$ (first term of (\ref{eqn:set2_III})). Therefore, it is possible to decode $m_2$ first. The second level for the strong interference condition at Rx1, assumes that after decoding $m_2$, the cognitive message ($m_3$) causes strong interference in comparison to the desired message ($m_1$) (first term of (\ref{eqn:set1_III})).
\end{itemize}
The above intuitions are summarized in Table~\ref{tbl:str_cond}.

\begin{table*}
\caption{Strong interference conditions}
\label{tbl:str_cond} \centering
\begin{tabular}{|c|c|c|c|c|c|} \hline
 & Tx3 $\rightarrow$ Rx1  & Tx3 $\rightarrow$ Rx2  &  Tx1 $\rightarrow$ Rx2&  Tx2 $\rightarrow$ Rx1&  Tx1 and Tx2 $\rightarrow$ Rx3\\
\hline \hline
\textit{Set1} & first term of (\ref{eqn:set1_I}) & second term of (\ref{eqn:set1_I}) & first term of (\ref{eqn:set1_II}) & first term of (\ref{eqn:set1_III}) & (\ref{eqn:set1_IV}) + second terms of (\ref{eqn:set1_II}) and (\ref{eqn:set1_III})\\
\hline
\textit{Set2} & first term of (\ref{eqn:set1_I}) & second term of (\ref{eqn:set1_I}) & first term of (\ref{eqn:set2_II}) & first term of (\ref{eqn:set2_III}) & (\ref{eqn:set2_IV}) + second terms of (\ref{eqn:set2_II}) and (\ref{eqn:set2_III})\\
\hline
\end{tabular}
\end{table*}

\begin{remark}
Theorem~\ref{thm:ach2} includes (\ref{eqn:ach2_I})-(\ref{eqn:ach2_III}) instead of (\ref{eqn:ach1_IV}) in Theorem~\ref{thm:ach1}. In fact, in the \emph{Gaussian case}, the converse proof can not be established for the two first terms in (\ref{eqn:ach1_IV}). Therefore, we propose Theorem~\ref{thm:ach2} and find the stronger conditions than \textit{Set1}, i.e., \textit{Set2}, which makes the bounds in (\ref{eqn:ach2_I})-(\ref{eqn:ach2_III}) redundant. Hence, we intend to use \textit{Set2} to derive the capacity results for the Gaussian case in Section~\ref{sec:Gaussian}.

In \textit{Set1}, (\ref{eqn:set1_IV}) and the second terms of (\ref{eqn:set1_II}) and (\ref{eqn:set1_III}) are used to make the second terms of (\ref{eqn:ach1_II}) and (\ref{eqn:ach1_III}), and the third term of (\ref{eqn:ach1_IV}) redundant. However, (\ref{eqn:set1_I}) and the first terms of (\ref{eqn:set1_II}) and (\ref{eqn:set1_III}) are used to prove the converse part for the rates in (\ref{eqn:Cap1_II})-(\ref{eqn:Cap1_IV}).

In \textit{Set2}, the second terms of (\ref{eqn:set2_II}) and (\ref{eqn:set2_III}) are used to make the second terms of (\ref{eqn:ach1_II}) and (\ref{eqn:ach1_III}) redundant.  The first terms of (\ref{eqn:set2_II}) and (\ref{eqn:set2_III}) make the (\ref{eqn:ach2_I}) and (\ref{eqn:ach2_II}) redundant and (\ref{eqn:set2_IV}) is used to make the (\ref{eqn:ach2_III}) redundant. However, (\ref{eqn:set1_I}) is used to prove the converse part for the rates in (\ref{eqn:Cap2_II}) and (\ref{eqn:Cap2_III}).

These results are summarized in Table~\ref{tbl:str_cond2}.
\end{remark}

\begin{table*}
\caption{Use of strong interference conditions in the achievability and converse proofs}
\label{tbl:str_cond2} \centering
\begin{tabular}{|c|c|c|} \hline
 & Achievability & Converse\\
\hline \hline
\textit{Set1} & (\ref{eqn:set1_IV}) + second terms of (\ref{eqn:set1_II}),(\ref{eqn:set1_III})  & (\ref{eqn:set1_I}) $\rightarrow$ (\ref{eqn:Cap1_II}),(\ref{eqn:Cap1_III})\\
& $\rightarrow$ second terms of (\ref{eqn:ach1_II}),(\ref{eqn:ach1_III}) + third term of (\ref{eqn:ach1_IV}) & (\ref{eqn:set1_I}) + first terms of (\ref{eqn:set1_II}),(\ref{eqn:set1_III}) $\rightarrow$ (\ref{eqn:Cap1_IV})\\
\hline
\textit{Set2} & second terms of (\ref{eqn:set2_II}),(\ref{eqn:set2_III}) $\rightarrow$ (\ref{eqn:ach1_II}),(\ref{eqn:ach1_III})& (\ref{eqn:set1_I}) $\rightarrow$ (\ref{eqn:Cap2_II}),(\ref{eqn:Cap2_III})\\
&first terms of (\ref{eqn:set2_II}),(\ref{eqn:set2_III}) $\rightarrow$ (\ref{eqn:ach2_I}),(\ref{eqn:ach2_II}) & \\
& (\ref{eqn:set2_IV}) $\rightarrow$ (\ref{eqn:ach2_III})&\\
\hline
\end{tabular}
\end{table*}

Assume that the following set of strong interference conditions (\textit{Set1}) holds for every $p(.)\in\Pc$:
\begin{IEEEeqnarray}{rcl}
I(X_3;Y_3|X_1,X_2)&\leq& \min\{I(X_3;Y_1|X_1,X_2),I(X_3;Y_2|X_1,X_2)\}\label{eqn:set1_I}\\
I(X_1,X_3;Y_1|X_2)&\leq& \min\{I(X_1;Y_2|X_2),I(X_1,X_3;Y_3|X_2)\}\label{eqn:set1_II}\\
I(X_2,X_3;Y_2|X_1)&\leq& \min\{I(X_2;Y_1|X_1),I(X_2,X_3;Y_3|X_1)\}\label{eqn:set1_III}\\
\min\{I(X_1,X_2,X_3;Y_1),I(X_1,X_2,X_3;Y_2)\}&\leq&I(X_1,X_2,X_3;Y_3) \label{eqn:set1_IV}
\end{IEEEeqnarray}
In fact, under these conditions interfering signals at the receivers are strong enough that all messages can be jointly decoded by all the receivers.
\begin{theorem}\label{thm:cap_set1}
The capacity region of the three-user C-IFC, satisfying (\ref{eqn:set1_I})-(\ref{eqn:set1_IV}), is given by
\begin{IEEEeqnarray}{rl}
\Cc_1 =\bigcup\limits_{p(.)\in\Pc} \big\{(R_1,R_2,R_3): &R_1 \geq 0, R_2 \geq 0, R_3 \geq 0 \nonumber\\
&R_{3}\leq I(X_3;Y_3|X_1,X_2) \label{eqn:Cap1_I}\\
&R_{1}+R_{3}\leq I(X_1,X_3;Y_1|X_2) \label{eqn:Cap1_II}\\
&R_{2}+R_{3}\leq I(X_2,X_3;Y_2|X_1) \label{eqn:Cap1_III}\\
&R_{1}+R_{2}+R_{3}\!\!\leq\!\! \min\{I(X_1,X_2,X_3;Y_1),I(X_1,X_2,X_3;Y_2)\}\big\}. \label{eqn:Cap1_IV}
\end{IEEEeqnarray}
\end{theorem}
\begin{remark}\label{remark:cap_set1_I}
The message of the cognitive user ($M_3$) can be decoded at Rx1 and Rx2, under condition (\ref{eqn:set1_I}). Rx1 can decode $M_2$ considering the condition of the first term in the RHS of (\ref{eqn:set1_III}). Note that, $X_3$ is required in this condition due to the dependance on $M_2$. Similarly, the condition of the first term in the RHS of (\ref{eqn:set1_II}) enables Rx2 to decode $M_1$. Moreover, $(M_1,M_2)$ can be decoded at Rx3 under (\ref{eqn:set1_IV}) and the second terms in the RHS of (\ref{eqn:set1_II}) and (\ref{eqn:set1_III}). Therefore, $\Cc_1$ gives the capacity region for a compound three-user MAC with common information, where $R_1$ and $R_2$ are the common rates of Tx1-Tx3 and Tx2-Tx3, respectively, $R_3$ is the private rate for Tx3, and the private rates for Tx1 and Tx2 are zero.
\end{remark}

\begin{remark}\label{remark:cap_set1_II}
If we omit the second pair, i.e., $X_2=Y_2=\emptyset$ and $R_2=0$, the model reduces to a two-user C-IFC and $\Cc_1$ coincides with the capacity region of the strong interference channel with unidirectional cooperation, which was characterized in \cite{MariYatKra07}.
\end{remark}

First, we provide a useful lemma which we need in the proof of the converse part for Theorem~\ref{thm:cap_set1}.

\begin{lemma}\label{lemma:cond_str^n}
If (\ref{eqn:set1_I})-(\ref{eqn:set1_III}) hold for all distribution $p(.)\in\Pc$, then we have:
\begin{IEEEeqnarray}{rcl}
I(X_3^n;Y_3^n|X_1^n,X_2^n,U)&\leq&I(X_3^n;Y_1^n|X_1^n,X_2^n,U)\label{eqn:set1_I^n1}\\
I(X_3^n;Y_3^n|X_1^n,X_2^n,U)&\leq&I(X_3^n;Y_2^n|X_1^n,X_2^n,U)\label{eqn:set1_I^n2}\\
I(X_1^n,X_3^n;Y_1^n|X_2^n,U)&\leq&I(X_1^n;Y_2^n|X_2^n,U)\label{eqn:set1_II^n}\\
I(X_2^n,X_3^n;Y_2^n|X_1^n,U)&\leq&I(X_2^n;Y_1^n|X_1^n,U).\label{eqn:set1_III^n}
\end{IEEEeqnarray}
\end{lemma}
\begin{IEEEproof}
Proof relies on the result in \cite[Proposition 1]{KorMar77} and follows the same lines as in \cite[Lemma~5]{MariYatKra07} and \cite[Lemma]{CosElg79}.
\end{IEEEproof}

\begin{IEEEproof}[Proof of Theorem~\ref{thm:cap_set1}]

\underline{Achievability:} Considering (\ref{eqn:set1_II})-(\ref{eqn:set1_IV}), the proof follows from Theorem~\ref{thm:ach1}.

\underline{Converse:} Consider a $(2^{nR_1},2^{nR_2},2^{nR_3},n)$ code with average error probability $P_e^n\rightarrow 0$, which implies that $P_{e,u}^{(n)}\rightarrow 0$ for $u\in\{1,2,3\}$. Applying Fano's inequality \cite{CovTho06}, \cite[P.~19]{ElgKim11} results in
\begin{IEEEeqnarray}{rcl}
H(M_u|Y_u^n)\leq P_{e,u}^{(n)} log(2^{nR_u} - 1) + h(P_{e,u}^{(n)})\leq n\delta_{un}\label{eqn:Fano_delta}
\end{IEEEeqnarray}
for $u\in\{1,2,3\}$, where $\delta_{un}\rightarrow 0$ as $P_{e,u}^{(n)}\rightarrow 0$. Note that, due to the encoding functions $f_1$, $f_2$ and $f_3$, defined in Definition~\ref{def:code} and the independence of the messages, we have $p(.)\in\Pc$. Now, we derive the bounds in Theorem~\ref{thm:cap_set1}. For the first bound, we obtain
\begin{IEEEeqnarray*}{ll}
nR_3=H(M_3)&\stackrel{(a)}{=}H(M_3|M_1,M_2)\\
&=I(M_3;Y_3^n|M_1,M_2)+H(M_3|Y_3^n,M_1,M_2)\\
&\stackrel{(b)}{\leq}I(M_3;Y_3^n|M_1,M_2)+H(M_3|Y_3^n)\yesnumber\label{eqn:cap_set1_fanoI_before}\\
&\stackrel{(c)}{\leq}I(M_3;Y_3^n|M_1,M_2)+n\delta_{3n}
\end{IEEEeqnarray*}
where (a) follows since $M_1$, $M_2$ and $M_3$ are independent, (b) is due to the fact that conditioning does not increase the entropy and (c) follows from (\ref{eqn:Fano_delta}) for $u=3$. Hence,
\begin{IEEEeqnarray*}{ll}
nR_3- n\delta_{3n}&\leq I(M_3;Y_3^n|M_1,M_2)\\
&\stackrel{(a)}{=} I(M_3,X_3^n;Y_3^n|M_1,M_2,X_1^n,X_2^n)\\
&=H(Y_3^n|M_1,M_2,X_1^n,X_2^n)-H(Y_3^n|M_1,M_2,X_1^n,X_2^n,M_3,X_3^n)\\
&\stackrel{(b)}{\leq} H(Y_3^n|X_1^n,X_2^n)-H(Y_3^n|M_1,M_2,X_1^n,X_2^n,M_3,X_3^n)\\
&\stackrel{(c)}{=} H(Y_3^n|X_1^n,X_2^n)-H(Y_3^n|X_1^n,X_2^n,X_3^n)=I(X_3^n;Y_3^n|X_1^n,X_2^n)\\
&\stackrel{(d)}{=}\sum\limits_{i=1}^{n}I(X_3^n;Y_{3,i}|X_1^n,X_2^n,Y_3^{i-1})\yesnumber\label{eqn:cap_set1_fanoI}\\
&=\sum\limits_{i=1}^{n}H(Y_{3,i}|X_1^n,X_2^n,Y_3^{i-1})-I(Y_{3,i}|X_1^n,X_2^n,Y_3^{i-1},X_3^n)\\
&\stackrel{(e)}{\leq}\sum\limits_{i=1}^{n}H(Y_{3,i}|X_{1,i},X_{2,i})-I(Y_{3,i}|X_1^n,X_2^n,Y_3^{i-1},X_3^n)\\
&\stackrel{(f)}{=}\sum\limits_{i=1}^{n}H(Y_{3,i}|X_{1,i},X_{2,i})-I(Y_{3,i}|X_{1,i},X_{2,i},X_{3,i})=\sum\limits_{i=1}^{n}I(X_{3,i};Y_{3,i}|X_{1,i},X_{2,i})
\end{IEEEeqnarray*}where (a) is due to the encoding functions $f_1$, $f_2$ and $f_3$, defined in Definition~\ref{def:code}, (b) and (e) are due to the fact that conditioning does not increase the entropy, (c) follows from the fact that $(M_1,M_2,M_3)\rightarrow (X_1^n,X_2^n,X_3^n)\rightarrow Y_3^n$ forms a Markov chain, (d) is obtained from the chain rule, and (f) follows from the memoryless property of the channel.

Now, applying (\ref{eqn:Fano_delta}) for $u\in\{1,3\}$ and the independence of the messages, we can bound $R_1+R_3$ as
\begin{IEEEeqnarray*}{ll}
n(R_1+R_3)- n(\delta_{1n}+\delta_{3n})&\leq I(M_1;Y_1^n|M_2)+I(M_3;Y_3^n|M_1,M_2)\\
&\stackrel{(a)}{=} I(M_1,X_1^n;Y_1^n|M_2,X_2^n)+I(M_3,X_3^n;Y_3^n|M_1,M_2,X_1^n,X_2^n)\\
&\stackrel{(b)}{=} I(M_1,X_1^n;Y_1^n|M_2,X_2^n)+I(X_3^n;Y_3^n|M_1,M_2,X_1^n,X_2^n)\\
&\stackrel{(c)}{\leq} I(M_1,X_1^n;Y_1^n|M_2,X_2^n)+I(X_3^n;Y_1^n|M_1,M_2,X_1^n,X_2^n)\\
&= I(M_1,X_1^n,X_3^n;Y_1^n|M_2,X_2^n)\yesnumber\label{eqn:cap_set1_fanoII}\\
&\stackrel{(d)}{=}\sum\limits_{i=1}^{n}I(M_1,X_1^n,X_3^n;Y_{1,i}|M_2,X_2^n,Y_1^{i-1})\\
&\stackrel{(e)}{\leq}\sum\limits_{i=1}^{n}I(X_{1,i},X_{3,i};Y_{1,i}|X_{2,i})
\end{IEEEeqnarray*}
where (a) follows from the encoding functions $f_1$, $f_2$ and $f_3$, defined in Definition~\ref{def:code}, (b) follows from the fact that $M_3\rightarrow (X_1^n,X_2^n,X_3^n)\rightarrow Y_3^n$ forms a Markov chain, (c) is obtained from (\ref{eqn:set1_I^n1}), (d) follows from the chain rule, and (e) follows from the memoryless property of the channel and the fact that conditioning does not increase the entropy (like parts (d)-(f) in (\ref{eqn:cap_set1_fanoI})). Applying similar steps using (\ref{eqn:Fano_delta}) for $u\in\{2,3\}$ and (\ref{eqn:set1_I^n2}), we can show that,
\begin{IEEEeqnarray}{l}
n(R_2+R_3)-n(\delta_{2n}+\delta_{3n}){\leq}\sum\limits_{i=1}^{n}I(X_{2,i},X_{3,i};Y_{2,i}|X_{1,i}).\label{eqn:cap_set1_fanoIII}
\end{IEEEeqnarray}

Finally, using (\ref{eqn:Fano_delta}) for $u\in\{1,2,3\}$ and the independence of the messages, the sum-rate bounds can be obtained as
\begin{IEEEeqnarray*}{rl}
n(R_1+R_2+R_3)- &n(\delta_{1n}+\delta_{2n}+\delta_{3n}){\leq} I(M_1;Y_1^n)+I(M_2;Y_2^n|M_1,M_3)+I(M_3;Y_3^n|M_1,M_2)\\
\stackrel{(a)}{\leq}&I(M_1,X_1^n;Y_1^n)+I(M_2,M_3,X_2^n,X_3^n;Y_2^n|M_1,X_1^n)\\
&+I(M_3,X_3^n;Y_3^n|M_1,M_2,X_1^n,X_2^n)\\
=&I(M_1,X_1^n;Y_1^n)+I(M_2,M_3,X_2^n,X_3^n;Y_2^n|M_1,X_1^n)\\
&+H(Y_3^n|M_1,M_2,X_1^n,X_2^n)-H(Y_3^n|M_1,M_2,X_1^n,X_2^n,M_3,X_3^n)\\
\stackrel{(b)}{\leq}&I(M_1,X_1^n;Y_1^n)+I(M_2,M_3,X_2^n,X_3^n;Y_2^n|M_1,X_1^n)\\
&+H(Y_3^n|M_1,X_1^n,X_2^n)-H(Y_3^n|M_1,M_2,X_1^n,X_2^n,M_3,X_3^n)\\
\stackrel{(c)}{\leq}&I(M_1,X_1^n;Y_1^n)+I(X_2^n,X_3^n;Y_2^n|M_1,X_1^n)+I(X_3^n;Y_3^n|M_1,X_1^n,X_2^n)\\
\stackrel{(d)}{\leq}&I(M_1,X_1^n;Y_1^n)+I(X_2^n;Y_1^n|M_1,X_1^n)+I(X_3^n;Y_1^n|M_1,X_1^n,X_2^n)\\
= &I(M_1,X_1^n,X_2^n,X_3^n;Y_1^n)\\
\stackrel{(e)}{=}&\sum\limits_{i=1}^{n}I(M_1,X_1^n,X_2^n,X_3^n;Y_{1,i}|Y_1^{i-1})\\
\stackrel{(f)}{\leq}&\sum\limits_{i=1}^{n}I(X_{1,i},X_{2,i},X_{3,i};Y_{1,i})\yesnumber\label{eqn:cap_set1_fanoIV}
\end{IEEEeqnarray*}
where (a) follows from the encoding functions $f_1$, $f_2$ and $f_3$, defined in Definition~\ref{def:code}, and the fact that conditioning does not increase the entropy, (b) is due to the fact that conditioning does not increase the entropy, (c) follows since $(M_2,M_3)\rightarrow (X_1^n,X_2^n,X_3^n)\rightarrow (Y_2^n,Y_3^n)$ forms a Markov chain, (d) is obtained from (\ref{eqn:set1_I^n1}) and (\ref{eqn:set1_III^n}), (e) follows from the chain rule, and (f) is due to the memoryless property of the channel and the fact that conditioning does not increase the entropy (like parts (d)-(f) in (\ref{eqn:cap_set1_fanoI})).

By applying a similar technique based on (\ref{eqn:set1_I^n2}) and (\ref{eqn:set1_II^n}), we obtain:
\begin{IEEEeqnarray*}{rcl}
n(R_1+R_2+R_3)&-& n(\delta_{1n}+\delta_{2n}+\delta_{3n}){\leq}\sum\limits_{i=1}^{n}I(X_{1,i},X_{2,i},X_{3,i};Y_{2,i}).\yesnumber\label{eqn:cap_set1_fanoV}
\end{IEEEeqnarray*}

Using a standard time-sharing argument \cite[P.~85]{ElgKim11} for (\ref{eqn:cap_set1_fanoI})-(\ref{eqn:cap_set1_fanoV}) completes the proof.
\end{IEEEproof}

Next, we derive the second set of strong interference conditions (\textit{Set2}), under which the region of Theorem~\ref{thm:ach2} is the capacity region. For every $p(.)\in\Pc$, \textit{Set2} includes (\ref{eqn:set1_I}) and the following conditions:
\begin{IEEEeqnarray}{rcl}
I(X_1,X_3;Y_1|X_2)&\leq& \min\{I(X_1;Y_2),I(X_1,X_3;Y_3|X_2)\}\label{eqn:set2_II}\\
I(X_2,X_3;Y_2|X_1)&\leq& \min\{I(X_2;Y_1),I(X_2,X_3;Y_3|X_1)\}\label{eqn:set2_III}\\
\{I(X_1;Y_2)\leq I(X_1;Y_3)\}&\textrm{ or }&\{I(X_2;Y_1)\leq I(X_2;Y_3)\} \label{eqn:set2_IV}
\end{IEEEeqnarray}

\begin{theorem}\label{thm:cap_set2}
The capacity region of the three-user C-IFC, satisfying  (\ref{eqn:set1_I}) and (\ref{eqn:set2_II})-(\ref{eqn:set2_IV}), is given by
\begin{IEEEeqnarray}{rl}
\Cc_2 =\bigcup\limits_{p(.)\in\Pc} \big\{(R_1,R_2,R_3): &R_1 \geq 0, R_2 \geq 0, R_3 \geq 0 \nonumber\\
&R_{3}\leq I(X_3;Y_3|X_1,X_2) \label{eqn:Cap2_I}\\
&R_{1}+R_{3}\leq I(X_1,X_3;Y_1|X_2) \label{eqn:Cap2_II}\\
&R_{2}+R_{3}\leq I(X_2,X_3;Y_2|X_1) \label{eqn:Cap2_III}\}\big\}.
\end{IEEEeqnarray}
\end{theorem}
\begin{IEEEproof}
\underline{Achievability:} Consider the region of Theorem~\ref{thm:ach2}. Using the second terms of conditions (\ref{eqn:set2_II}) and (\ref{eqn:set2_III}), the bounds in (\ref{eqn:ach1_I})-(\ref{eqn:ach1_III}) reduce to $\Cc_2$. Based on the first term of condition (\ref{eqn:set2_II}), the bound in (\ref{eqn:ach2_I}) is redundant due to (\ref{eqn:Cap1_II}). Similarly, (\ref{eqn:Cap1_III}) and (\ref{eqn:set2_III}) make the bound in (\ref{eqn:ach2_II}) redundant. Moreover, considering (\ref{eqn:ach2_I}) (or (\ref{eqn:ach2_II})), (\ref{eqn:set2_IV}), and the second bound in (\ref{eqn:ach1_III}) (or (\ref{eqn:ach1_II})), the bound in (\ref{eqn:ach2_III}) becomes redundant.

\underline{Converse:} The bounds in $\Cc_2$ are same as the bounds (\ref{eqn:Cap1_I})-(\ref{eqn:Cap1_III}) in $\Cc_1$, which are shown in the converse proof of Theorem~\ref{thm:cap_set1}. This completes the proof.
\end{IEEEproof}

\begin{remark}[Comparison of two sets of conditions]
We compare the different terms in \textit{Set1} and \textit{Set2}. Since $X_1$ and $X_2$ are independent, we obtain
\begin{IEEEeqnarray*}{rl}
I(X_2;Y_1|X_1)=H(X_2|X_1)-H(X_2|X_1,Y_1)&\stackrel{(a)}{=}H(X_2)-H(X_2|X_1,Y_1)\\
&\stackrel{(b)}{\geq}H(X_2)-H(X_2|Y_1)= I(X_2;Y_1)
\end{IEEEeqnarray*}
where (a) follows from the independence of $X_1$ and $X_2$, and (b) is a consequence of the fact that conditioning does not increase the entropy.
Hence, condition (\ref{eqn:set2_III}) implies condition (\ref{eqn:set1_III}). Similarly, condition (\ref{eqn:set2_II}) implies condition (\ref{eqn:set1_II}). Moreover, the second terms of conditions (\ref{eqn:set2_II}) and (\ref{eqn:set2_IV}) give the first term in condition (\ref{eqn:set1_IV}). Also, the second term of condition (\ref{eqn:set2_III}) along with the first term of (\ref{eqn:set2_IV}) give the second term in condition (\ref{eqn:set1_IV}). Therefore, \textit{Set2} implies \textit{Set1}, and the conditions of \textit{Set1} are weaker compared to thoes of \textit{Set2}. In fact, we use \textit{Set2} and $\Cc_2$ to derive the capacity results for the Gaussian case in the next section.
\end{remark}

\section{Gaussian three-user C-IFC}\label{sec:Gaussian}
\begin{figure}[tb]
  \centering
  \includegraphics[width=9cm]{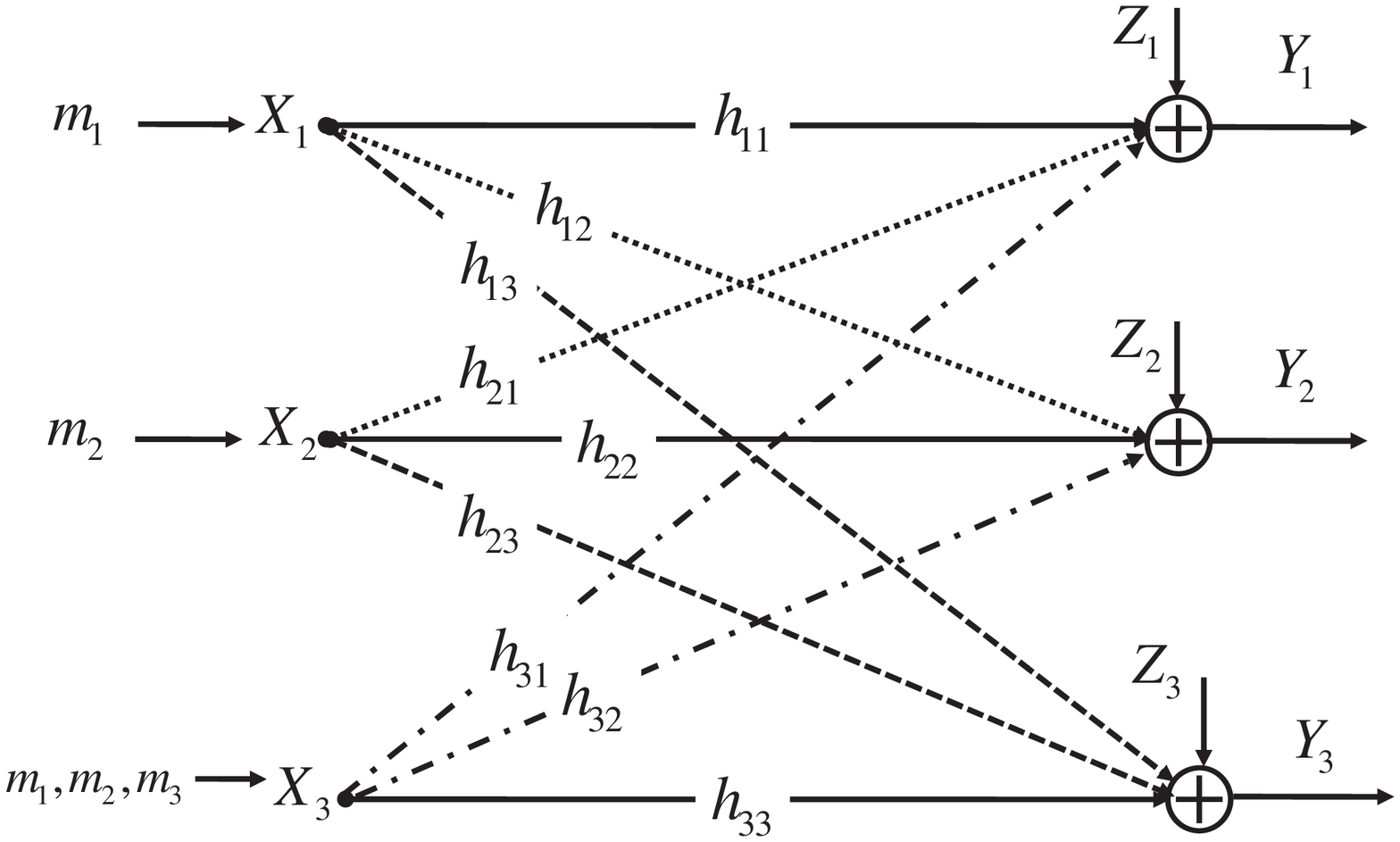}
  \caption{Gaussian three-user C-IFC.}
  \label{fig:Gauschannelmodel}
\end{figure}

In this section, we consider the Gaussian three-user C-IFC and characterize capacity results for the Gaussian case. Moreover, we present some numerical examples. 
The Gaussian three-user C-IFC, as depicted in Fig.~\ref{fig:Gauschannelmodel}, at time $i=1,\ldots,n$ and at each Rx$r$, for $r\in\{1,2,3\}$, can be mathematically modeled as
\begin{IEEEeqnarray}{rcl}
Y_{r,i}&\:=\:&\sum\limits_{t=1}^{3}h_{tr}X_{t,i}+Z_{r,i}\label{eqn:Gaussian_model}
\end{IEEEeqnarray}
where $h_{tr}$, for $t,r\in\{1,2,3\}$, are known channel gains. $X_{t,i}$ is the input signal with average power constraint: 
\begin{IEEEeqnarray}{rcl}
\frac{1}{n}\sum\limits_{i=1}^n(x_{t,i})^2\leq P_t\label{eqn:power_cons}
\end{IEEEeqnarray}
for $t\in\{1,2,3\}$. $Z_{r,i},\: r\in\{1,2,3\}$ is an independent and identically distributed (i.i.d) zero mean Gaussian noise component with unit power, i.e., $Z_{r,i}\sim\Nc(0,1)$.

Now, we extend the results of Theorem~\ref{thm:cap_set2}, i.e., $\Cc_2$ and \textit{Set2}, to the Gaussian case. The strong interference conditions of \textit{Set2}, i.e., (\ref{eqn:set1_I}), (\ref{eqn:set2_II})-(\ref{eqn:set2_IV}) for the above Gaussian model, respectively, become ($Set_G$):
\begin{IEEEeqnarray*}{rcl}
h_{33}^2&\leq&\min\{h_{31}^2,h_{32}^2\}\yesnumber\label{eqn:set1_I_Gaus}\\
h_{11}^2P_1+h_{31}^2P_3(1-\rho_2^2)+2h_{11}h_{31}\rho_1\sqrt{P_1P_3}&\leq&\min\big\{A_{12},h_{13}^2P_1+h_{33}^2P_3(1-\rho_2^2)+2h_{13}h_{33}\rho_1\sqrt{P_1P_3}\big\}\\
&&\yesnumber\label{eqn:set1_II_Gaus}\\
h_{22}^2P_2+h_{32}^2P_3(1-\rho_1^2)+2h_{22}h_{32}\rho_2\sqrt{P_2P_3}&\leq&\min\big\{A_{21},h_{23}^2P_1+h_{33}^2P_3(1-\rho_1^2)+2h_{23}h_{33}\rho_2\sqrt{P_2P_3}\big\}\\
&&\yesnumber\label{eqn:set1_III_Gaus}\\
\{A_{12}\leq B_{12}\}&\textrm{ or }&\{A_{21}\leq B_{21}\}\yesnumber\label{eqn:set1_IV_Gaus}
\end{IEEEeqnarray*}
where $-1\leq\rho_u\leq 1$ is the correlation coefficient between $X_u$ and $X_3$, i.e., $E(X_uX_3)=\rho_u\sqrt{P_uP_3}$ for $u\in\{1,2\}$, and $A_{ij}$ and $B_{ij}$ are defined as,
\begin{IEEEeqnarray*}{rcl}
A_{ij}&=&\frac{(h_{ij}\sqrt{P_i}+h_{3j}\rho_i\sqrt{P_3})^2}{h_{jj}^2P_j+h_{3j}^2P_3(1-\rho_i^2)+2h_{jj}h_{3j}\rho_j\sqrt{P_jP_3}+1}\\ B_{ij}&=&\frac{(h_{i3}\sqrt{P_i}+h_{33}\rho_i\sqrt{P_3})^2}{h_{j3}^2P_j+h_{33}^2P_3(1-\rho_i^2)+2h_{j3}h_{33}\rho_j\sqrt{P_jP_3}+1}
\end{IEEEeqnarray*}
for $i,j\in \{1,2\}$.

\begin{theorem}\label{thm:Gaus_cap_set1}
For the Gaussian three-user C-IFC, satisfying conditions (\ref{eqn:set1_I_Gaus})-(\ref{eqn:set1_IV_Gaus}), the capacity region is given by
\begin{IEEEeqnarray}{rl}
\Cc_1^G =\bigcup\limits_{-1\leq\rho_1,\rho_2\leq 1:\rho_1^2+\rho_2^2\leq 1 } \big\{&(R_1,R_2,R_3): R_1,R_2,R_3 \geq 0 \nonumber\\
&R_3\leq \theta(h_{33}^2P_3(1-\rho_1^2-\rho_2^2))\label{eqn:Gaus_Cap1_I}\\
&R_{1}+R_{3}\leq \theta(h_{11}^2P_1+h_{31}^2P_3(1-\rho_2^2)+2h_{11}h_{31}\rho_1\sqrt{P_1P_3})\label{eqn:Gaus_Cap1_II}\\
&R_{2}+R_{3}\leq \theta(h_{22}^2P_2+h_{32}^2P_3(1-\rho_1^2)+2h_{22}h_{32}\rho_2\sqrt{P_2P_3})\!\big\}\label{eqn:Gaus_Cap1_III}
\end{IEEEeqnarray}
where to simplify notation we define
\begin{equation}\label{eqn:theta}
\theta(x)\doteq \frac{1}{2}\log(1+x).
\end{equation}
\end{theorem}

\begin{remark}\label{remark:Gaus_cap_set1_I}
Condition (\ref{eqn:set1_I_Gaus}) implies that Tx3 causes strong interference at Rx1 and Rx2. This fact enables Rx1 and Rx2 to decode $m_3$. Moreover, due to the first terms in the RHS of (\ref{eqn:set1_II_Gaus}) and (\ref{eqn:set1_III_Gaus}), $m_1$ and $m_2$ can be decoded at Rx2 and Rx1, respectively. Also, (\ref{eqn:set1_II_Gaus})-(\ref{eqn:set1_IV_Gaus}) provides strong interference conditions at Rx3, under which all messages can be decoded by Rx3.
\end{remark}

\begin{IEEEproof}
The achievablity follows from $\Cc_2$ in Theorem~\ref{thm:cap_set2} by evaluating \textit{Set2} and $\Cc_2$ with zero mean jointly Gaussian channel inputs $X_1$, $X_2$ and $X_3$. In fact, $X_1\sim\Nc(0,P_1)$, $X_2\sim\Nc(0,P_2)$ and $X_3\sim\Nc(0,P_3)$, where $E(X_1X_2)=0$, $E(X_1X_3)=\rho_1\sqrt{P_1P_3}$, and $E(X_2X_3)=\rho_2\sqrt{P_2P_3}$. The converse proof is based on the similar reasoning as in \cite{Sato81} and is provided in Appendix~\ref{app:thm:Gaus_converse}.
\end{IEEEproof}

Note that, the channel parameters, i.e., $h_{tr}$, $P_t$ for $t,r\in\{1,2,3\}$, must satisfy (\ref{eqn:set1_I_Gaus})-(\ref{eqn:set1_IV_Gaus}) for all $-1\leq\rho_1,\rho_2\leq 1:\rho_1^2+\rho_2^2\leq 1$, to numerically evaluate the $\Cc_1^G$ using (\ref{eqn:Gaus_Cap1_I})-(\ref{eqn:Gaus_Cap1_III}). Here, we choose $P_1=P_3=3$, $P_2=6$, $h_{11}=h_{22}=h_{33}=1$, $h_{31}=h_{32}=\sqrt{1.5}$, $h_{12}=7$, $h_{13}=3$, $h_{21}=5$, and $h_{23}=15$, which satisfy (\ref{eqn:set1_I_Gaus})-(\ref{eqn:set1_IV_Gaus}); hence, the regions are derived under strong interference conditions $Set_G$.

\begin{figure}[tb]
  \centering
  \includegraphics[width=11cm]{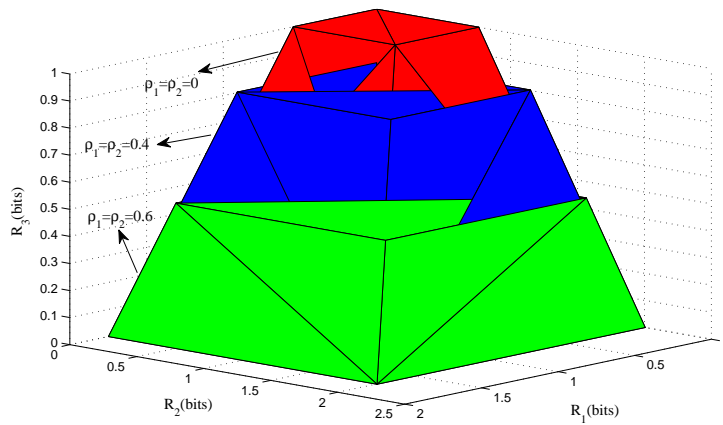}
  \caption{Capacity region for the Gaussian three-user C-IFC for fixed $\rho_1=\rho_2$.}
  \label{fig:Gaus1}
\end{figure}
\begin{figure}[tb]
  \centering
  \includegraphics[width=11cm]{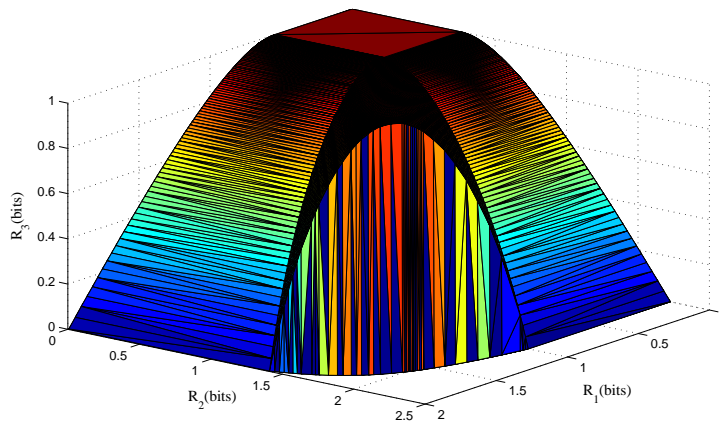}
  \caption{Capacity region for the Gaussian three-user C-IFC under strong interference conditions $Set_G$.}
  \label{fig:Gaus2}
\end{figure}

Fig.~\ref{fig:Gaus1} shows the capacity region for the Gaussian three-user C-IFC of Theorem~\ref{thm:Gaus_cap_set1}, for the above parameter selection, where $\rho_1=\rho_2$ is fixed in each surface. $\rho_1=\rho_2=0$ region corresponds to the \emph{no cooperation} case, where the channel inputs are independent. It can be seen that as $\rho_1=\rho_2$ increases, the bound on $R_3$ becomes more restrictive while the sum-rate bounds become looser; because Tx3 dedicates parts of its power for cooperation. The capacity for this channel is the union of all the regions obtained for different values of $\rho_1$ and $\rho_2$ satisfying $\rho_1^2+\rho_2^2\leq 1$. This union is shown in Fig.~\ref{fig:Gaus2}.

\section{Conclusion}\label{sec:conclusion}
We considered a three-user cognitive radio network with two primary users and one cognitive user and investigated its capacity region in the strong interference regime. For this purpose, we introduced the three-user Cognitive Interference Channel (C-IFC) by providing cognition capabilities for one of the transmitters in the three-user IFC. We derived two sets of strong interference conditions under which we established the capacity regions. Under these conditions, all three messages are required at all receivers. We also found capacity results for the Gaussian case.

\appendices
\section{Proofs of Theorem~\ref{thm:ach1} and Theorem~\ref{thm:ach2}}\label{app:ach_proof}

\begin{IEEEproof}[Outline of the proof of Theorem~\ref{thm:ach1}]
We propose the following random coding scheme, which contains superposition coding in the encoding part and simultaneous joint decoding in the decoding part. All messages are common to all receivers, i.e., all three receivers decode $m_1$, $m_2$ and $m_3$.

\textit{\textbf{Codebook Generation}:} Fix $p(.)\in\Pc$. For $u\in\{1,2\}$, generate $2^{nR_{u}}$ i.i.d $x_{u}^n$ sequences, each with probability
$\prod\limits_{i=1}^np(x_{u,i})$. Index them as $x_{u}^n(m_{u})$ where $m_{u}\in[1,2^{nR_{u}}]$. For each $(x_{1}^n(m_{1}),x_{2}^n(m_{2}))$, generate $2^{nR_{3}}$ i.i.d $x_{3}^n$ sequences, each with probability $\prod\limits_{i=1}^np(x_{3,i}|x_{1,i},x_{2,i})$. Index them as $x_{3}^n(m_{3},m_2,m_1)$ where $m_{3}\in[1,2^{nR_{3}}]$.

\textit{\textbf{Encoding}:} In order to transmit the message $(m_{1},m_{2},m_3)$, Txu sends $x_{u}^n(m_{u})$ for $u\in\{1,2\}$ and Tx3 sends $x_{3}^n(m_{3},m_2,m_1)$.

\textit{\textbf{Decoding}:}

\emph{Rx1:} After receiving $y_1^n$, Rx1 looks for a unique index $\hat{m}_{1}$ and some $(\hat{m}_{2},\hat{m}_{3})$ such that, \begin{IEEEeqnarray*}{c}
(y_1^n,x_{1}^n(\hat{m}_{1}),x_{2}^n(\hat{m}_{2}),x_{3}^n(\hat{m}_{3},\hat{m}_2,\hat{m}_1))\in A_\epsilon^n\left(Y_1,X_1,X_2,X_3\right).
\end{IEEEeqnarray*}

Using the packing lemma \cite[P.~45]{ElgKim11} (or \cite[Theorem 15.2.3]{CovTho06}), for large enough $n$, with arbitrarily high probability $\hat{m}_{1}=m_{1}$ if
\begin{IEEEeqnarray}{rcl}
R_{1}+R_{3}&\leq& I(X_1,X_3;Y_1|X_2)\label{eqn:ach1_rec1_I}\\
R_{1}+R_{2}+R_{3}&\leq& I(X_1,X_2,X_3;Y_1).\label{eqn:ach1_rec1_II}
\end{IEEEeqnarray}

\emph{Rx2:} The decoding process at Rx2 is similar to Rx1. Therefore, based on packing lemma \cite[P.~45]{ElgKim11}, the decoding error at Rx2 can be made sufficiently small if
\begin{IEEEeqnarray}{rcl}
R_{2}+R_{3}&\leq& I(X_2,X_3;Y_2|X_1)\label{eqn:ach1_rec2_I}\\
R_{1}+R_{2}+R_{3}&\leq& I(X_1,X_2,X_3;Y_2).\label{eqn:ach1_rec2_II}
\end{IEEEeqnarray}

\emph{Rx3:} After receiving $y_3^n$, Rx3 finds a unique index $\hat{\hat{m}}_{3}$ and some pair $(\hat{\hat{m}}_{1},\hat{\hat{m}}_{2})$ such that,
\begin{IEEEeqnarray*}{c}
(y_3^n,x_{1}^n(\hat{\hat{m}}_{1}),x_{2}^n(\hat{\hat{m}}_{2}),x_{3}^n(\hat{\hat{m}}_{3},\hat{\hat{m}}_2,\hat{\hat{m}}_1))\in A_\epsilon^n\left(Y_3,X_1,X_2,X_3\right).
\end{IEEEeqnarray*}

Using packing lemma \cite[P.~45]{ElgKim11}, With arbitrary high probability $\hat{\hat{m}}_{3}=m_3$, if $n$ is large enough and
\begin{IEEEeqnarray}{rcl}
R_{3}&\leq& I(X_3;Y_3|X_1,X_2) \label{eqn:ach1_rec3_I}\\
R_{1}+R_{3}&\leq& I(X_1,X_3;Y_3|X_2)\label{eqn:ach1_rec3_II}\\
R_{2}+R_{3}&\leq& I(X_2,X_3;Y_3 |X_1)\label{eqn:ach1_rec3_III}\\
R_{1}+R_{2}+R_{3}&\leq& I(X_1,X_2,X_3;Y_1).\label{eqn:ach1_rec3_IV}
\end{IEEEeqnarray}

This completes the proof.
\end{IEEEproof}

\begin{IEEEproof}[Outline of the proof of Theorem~\ref{thm:ach2}]
The \textit{codebook generation} and \textit{encoding} parts of the proof follow the same lines as in Theorem~\ref{thm:ach1}. Therefore, we only describe the decoding part. Similar to Theorem~\ref{thm:ach1}, all three receivers decode $m_1$, $m_2$ and $m_3$. However, here Rx1 (or Rx2) first decodes $m_2$ (or $m_1$). Then, it jointly decodes $m_1$ (or $m_2$) and $m_3$.

\textit{\textbf{Decoding}:}

\emph{Rx1:} After receiving $y_1^n$, Rx1 first finds a unique index $\hat{m}_{2}$ such that,
\begin{IEEEeqnarray*}{c}
(y_1^n,x_{2}^n(\hat{m}_{2}))\in A_\epsilon^n(Y_1,X_2).
\end{IEEEeqnarray*}

Applying packing lemma \cite[P.~45]{ElgKim11}, with arbitrary high probability $\hat{m}_{2}=m_{2}$, if $n$ is large enough and
\begin{IEEEeqnarray}{rcl}
R_{2}&\leq& I(X_2;Y_1). \label{eqn:ach2_rec1_I}
\end{IEEEeqnarray}

Then, it looks for a unique index $\hat{m}_{1}$ and some $\hat{m}_{3}$ such that,
\begin{IEEEeqnarray*}{c}
(y_1^n,x_{1}^n(\hat{m}_{1}),x_{2}^n(m_{2}),x_{3}^n(\hat{m}_{3},m_2,\hat{m}_1))\in A_\epsilon^n\left(Y_1,X_1,X_2,X_3\right).
\end{IEEEeqnarray*}

For large enough $n$, with arbitrarily high probability $\hat{m}_{1}=m_{1}$ if (\ref{eqn:ach1_rec1_I}) holds.

\emph{Rx2:} Rx2 proceeds similarly. This step can be accomplished with sufficiently small probability of error for large enough $n$, if (\ref{eqn:ach1_rec2_I}) holds and
\begin{IEEEeqnarray}{rcl}
R_{1}&\leq& I(X_1;Y_2).\label{eqn:ach2_rec2_I}
\end{IEEEeqnarray}

The decoding procedure at Rx3 is similar to Theorem~\ref{thm:ach1} and the error in this receiver can be bounded, if (\ref{eqn:ach1_rec3_I})-(\ref{eqn:ach1_rec3_IV}) hold. This completes the proof.
\end{IEEEproof}

\section{Proof of the Converse Part for Theorem~\ref{thm:Gaus_cap_set1}}\label{app:thm:Gaus_converse}
For any rate triple $(R_1,R_2,R_3)\in\Cc$, Rx1 is able to decode $m_1$ reliably. Assume that Rx1 knows $X_2$ by a genie. Obviously, the genie aided channel has a larger capacity region than $\Cc$. Now, Rx1 knows $X_1$ from $m_1$ and $X_2$ from genie. Then, Rx1 is able to construct
\begin{IEEEeqnarray*}{rcl}
\tilde{Y}_3&=&\frac{h_{33}}{h_{31}}(Y_1-h_{11}X_1-h_{12}X_2)+h_{13}X_1+h_{23}X_2\\
&=&h_{13}X_1+h_{23}X_2+h_{33}X_3+\frac{h_{33}}{h_{31}}Z_1
\end{IEEEeqnarray*}
If condition (\ref{eqn:set1_I_Gaus}) holds, then $\tilde{Y}_3$ is a less noisy version of $Y_3$. Since Rx3 has to decode $m_3$, Rx1 can decode $m_3$ via $\tilde{Y}_3$. Therefore, $(R_1,R_2,R_3)$ is contained in the capacity region of a MAC with common information from Tx1 and Tx3 to Rx1 with $X_2$ as a receiver side information, where $R_1$ is the common rate, $R_3$ is the private rate for Tx3, and the private rates for Tx1 is zero. Therefore, the sum-rate $R_1+R_3$ is bounded as (\ref{eqn:Cap2_II}). From the maximum-entropy theorem \cite{CovTho06} (or \cite[P.~21]{ElgKim11}), this bound is largest for the Gaussian inputs and is evaluated to (\ref{eqn:Gaus_Cap1_II}). In a similar manner, we can obtain (\ref{eqn:Gaus_Cap1_III}) at Rx2. The bound in (\ref{eqn:Gaus_Cap1_I}) follows by applying the standard methods as in (\ref{eqn:cap_set1_fanoI}).

%

%
%
%




\end{document}